\documentclass[12pt]{article}
 \usepackage{longtable}
\relax
\newcommand{\bo}{{\bar o}}

\def\bo{{\raise.15ex\hbox{\large$\Box$}}}               % D'Alembertian

\def\face{{\raise.2ex\hbox{$\displaystyle \bigodot$}\mskip-2.2mu \llap {$\ddot
        \smile$}}}                                      % happy face
\def\leftrightarrowfill{$\mathsurround=0pt \mathord\leftarrow \mkern-6mu
        \cleaders\hbox{$\mkern-2mu \mathord- \mkern-2mu$}\hfill
        \mkern-6mu \mathord\rightarrow$}       % <--> double differential
\def\dvec#1{\vbox{\ialign{##\crcr
        \leftrightarrowfill\crcr\noalign{\kern-1pt\nointerlineskip}
        $\hfil\displaystyle{#1}\hfil$\crcr}}}           % <--> accent
\def\beq{\begin{equation}}
\def\eeq{\end{equation}}

\def\beqx{\begin{displaymath}}
\def\eeqx{\end{displaymath}}

\def\beql{\begin{eqnarray}}
\def\eeql{\end{eqnarray}}

\newcommand{\bea}{\begin{eqnarray}}
\newcommand{\eea}{\end{eqnarray}}

\def\[{\left [}
\def\]{\right ]}
\def\({\left (}
\def\){\right )}

\def\+{\oplus}

\begin{document}

%\vspace*{0.15in}
%\hbox{\hskip 12cm NIKHEF/2010-....  \hfil}
\hbox{\hskip 12cm IFF/FM-2010/04  \hfil}
%\hbox{\hskip 12cm hep-th/yymmnnn \hfil}

\vskip .5in

\begin{center}
{\Large \bf  The Last Workshop on Theoretical Physics in the} 

\vskip .1in

{\Large \bf Soviet Union}

\vspace*{.7in}
{ B. Gato-Rivera}\footnote{Also known as B. Gato}
 
\vskip .5in

{\em Instituto de F\'\i sica Fundamental, CSIC, \\
Serrano 123, Madrid 28006, Spain} \\

\vskip .2in

\end{center}

\begin{center}
\vspace*{0.5in}
{\bf Abstract}
\end{center}

Twenty years ago, in October 1990, I found myself attending a workshop on Theoretical Physics 
in Chernomorka (Ukraine) intended only for Soviet physicists. That trip to the USSR/CCCP as well as the 
preceding months at CERN were highly surrealistic with plenty of adventures, crucial events and 
anecdotes, the most amazing one involving Niels Bohr. A few months later the Soviet Union 
collapsed. In this article I make a personal account on the happenings of 1990, with emphasis on
my incursions into the European communist world during, and also before, that year.

\vskip 1in

\noindent
December 2010\\
Revised version April 2011

\newpage

\noindent
{\bf The annus mirabilis}

\vskip .5in

1990 was some kind of `annus mirabilis' for me, both in the personal and professional sense.
Some important crucial events took place that year, as well as several other very special and
unique experiences, those that leave an imprint in one's life and one never forgets no matter
the time. First of all I got the permanent position of my dreams in Madrid in the CSIC (Spanish
National Research Council) starting the first of March. Moreover, I was allowed to stay at CERN
until the autumn 1992, since I had just arrived there in October 1989 after spending three
years as postdoc at the CTP (Center for Theoretical Physics) in MIT . This was a happy circumstance,
not only for professional reasons, but also because in the beginning of April I started going out 
with a colleague from CERN and we got married one year later. 

In the middle of April I had a very pleasant surprise: Julius Wess came to CERN to stay for two 
months. I had met Julius 
two years before, when he visited the CTP in MIT during four months. We got along very well from
the very beginning, talking to each other at a daily basis over a wide range of topics. (At the CTP 
the postdocs and visitors were invited to join the scientific staff for lunch in the common room,
which was also the seminar room. This was most convenient, of course, and allowed me to meet
Julius just after he arrived from Karlsruhe.) He was a very warm and kind person to everybody and also 
very sensitive to social matters in general\footnote{He was so sensitive that, just after the earthquake
in Istanbul in August 16th 1999, that we as participants of the Wigner Symposium lived through, he 
refused to accompany a numerous group of us for dinner to the restaurant of the famous Galatea tower, 
out of sorrow. It was impossible to help the victims, it was not even allowed to approach the sites, so a
large group of us decided to go out for dinner to one of the few open touristic places. We probably
needed to relax and have some fun. For example, I gave my talk only a few hours after the earthquake
with the ground vibrating continuously below my feet.}, and to 
discrimination of women in particular. He deeply regretted the lack of women in Physics and was 
very proud of Emmy Noether. His wife, Traudi, was also very friendly.  Shortly after they arrived 
they invited me for dinner at home, one of their daughters was also there, and during dinner we made 
plans to go for a one-day excursion to the mountains together with Bruno Zumino and Mary 
K. Gaillard who were also visiting CERN.

The excursion took place a few days later on a wonderful sunny day in La Sal\`eve (French Alps).
We walked up a mountain, about three or four hours walk, with Julius at the front with very much advantage
over the five of us (Bruno, Mary K., Traudi, my brand new partner and I). Traudi was last ..... because
she was pushing me on the back for at least two hours. Her help was crucial for my arrival at the top, as 
well as for my pride and honour. In fact,  I had never climbed a mountain before, and I have never tried to 
do it again, it was too exhausting. So, this was certainly a unique experience in my life (very charming, 
especially for the company and the landscape), although not to be repeated!  At the top all of us, except
Julius, were destroyed. We couldn't even talk while eating our lunches, with Julius making some of his typical 
innocent remarks regarding human nature. For example:  `All the badness and selfishness of people are 
due only to survival reasons'. I think he was the only one in the group who believed that\footnote{As 
a matter of fact, and contrary to Julius spirit, Traudi warned me a couple of times about the possible envy 
and jealousy of some wives of my colleagues with less professional status than me, who would try to 
influence their husbands against me. I must say that I don't know if she was right - I rarely have any 
contact with the wives of my colleagues -  but I am pretty sure that sexist colleagues don't need the 
help of their wives to be nasty towards their female colleagues (they can become even 
nastier, perhaps).}.

In May I attended the lectures on `Gravitation in curved spacetimes' by Ruth Williams. One of those
mornings just before the lecture John Bell, who was the organizer, told me I was invited for lunch in
a nice restaurant inside CERN that was used for special occasions. I understood immediately: this lunch
was in honour of Ruth, and John wanted to bring another woman at the table. So, I smiled at him with
complicity and said: `Thanks a lot, yes I would like to join you for lunch'. I had met John Bell shortly 
after I arrived at CERN, in October 1989. We were side by side behind computer terminals in the 
common room and somehow we started talking about how strange Quantum Mechanics was (and still is). 
That was in fact my only conversation with John until we went to the restaurant, so I was very excited
for having the opportunity to talk to him a bit more. 

At the table there were five British physicists and me
and my intuition was correct: Ruth and I were `the women'. The other three British guys were
Graham Shore, who was a `veteran' postdoc, and two very young ones (as usual with British 
postdocs). John was on my right side and we were facing Ruth. He ordered a vegetarian 
menu for himself. Most of the conversation was very standard: shortage of jobs in U. K. , etc.,
but at a given moment Ruth said something that ignited 
a lively discussion, especially between John and me. Ruth reported that the most important issue in
her life was her child. Immediately John and I reacted with some comments about how content we were
without children (I had been married for several years, in fact I had just divorced a couple of weeks 
ago, in April). Then John and I engaged in a conversation about this issue. I will never forget when he
explained to me that he and his wife Mary decided not to have children: `I was convinced that if we had
children Mary wouldn't be able to cope with her career (she was also physicist), so I decided not to have
children and she agreed that children wouldn't be convenient for her'. `But we have felt often pressure
from people who believe that the duty of all married couples is to have children. Do you think that we all
have the obligation to have children?' `No, of course not', I replied, `I actually think that only people very 
fond of children should have them, nobody else. In addition, at least half of the kids bring big troubles to
their parents sooner or later'. `Yes, exactly this has happened to my brothers and my friends....' John replied.
Etc, etc.

That conversation, together with the invitation to the restaurant, shows the high sensitivity of John Bell in
relation to women issues, as was also the case for Julius Wess.  Obviously that lunch was unforgettable 
for me, but also it was most unique because John Bell died four months later (October 1st) of a 
stroke after spending a few 
days in coma\footnote{I still remember as if it was yesterday when I asked Tatiana, the Russian Chief 
secretary of the Theory Division: `Tatiana, how is John Bell?', and she replied: `He is dead'.}.
   
1990 certainly looked promising: only in three months - from March to May -  I started one of the best jobs a 
scientist can get in Spain, I engaged in the personal relation with my present 
husband, I got divorced from the previous one, I climbed  `the Wess-Zumino' mountain and I was invited 
to a very charming unforgettable lunch by John Bell. The summer months were also very special and 
unique: traveling around CERN with friends from Madrid during two weeks in July and traveling all over
Switzerland with my fianc\'e during two weeks in August. We were driving from one 
wonderful scenario to another one even more wonderful, as if we were taking part in a fairy tale. I think 
I had never seen  so much natural beauty at once before, it was really beyond imagination. 

Back at CERN, I got an invitation to visit the Institute for Theoretical and Experimental Physics (ITEP)
in Moscow for two weeks during the autumn. It happened at that time that many Soviet physicists, 
mainly from ITEP, were invited to a small European tour, essentially to Paris and CERN, whereas a few
others were visiting CERN for a longer period of several months.
For most of them this was their first incursion into `the West', as they used to call the non-communist world.
Aliosha Morozov, who was the leader of the ITEP group, handed the invitation to me. I accepted gladly 
looking forward to visit the Soviet Union with interest and curiosity. I had been only once in the communist 
world, and for very short time, and expected to come back sometime in the future.

\vskip .5in

\noindent
{\bf Berlin, August 1979}

\vskip .5in

It was in Berlin in August 1979, during a cultural trip for foreign students of science and technology 
in the last two years in the university, who were working in Germany during the summer months 
under the interchange program IAESTE. I was working in the PTB (Physikalish-Technische 
Bundesanstalt) in Braunschweig searching for new lines of a $CO_2$  MASER 
in the laboratory of Carl Otto Weiss (one of the most charming scientists I have ever met).

The trip, one week in Berlin, included a cultural visit to the East part of the city as well as several 
instructive approaches to `the Wall', which was in fact two walls with nobody's land in between. 
The feeling was amazing: that white innocent looking wall (not even tall) was what it 
was. Of course, one could easily see appropriate decoration - reflectors, some objects that looked like 
machine guns, and nasty metal wire -  between the two walls. West Berlin seemed like a strange patchwork
to me with only three huge pieces: the French, the British and the American sectors. In the vicinity of 
the Wall there were many indications of the type: `Attention. You are leaving the American Sector', 
written in German and English, and the word Achtung was ubiquitous.

There were three access points between West Berlin and East Berlin : Checkpoint Charlie, 
the subway station Friedrichstrasse and the Brandenburger Tor (the Gate of Brandenburg), 
which was reserved only for buses. 
As long as our bus approached the Brandenburger Tor I became increasingly excited since I had 
seen the place in several postcards and now the real and beautiful monument would unfold in front of
my eyes. The access was located on the left of the gate (as seen from West Berlin), which
was the right side of the front of the monument as regarded front East Berlin.  We were
essentially alone there, no queue of buses at all\footnote{This, together 
with the fact that it was August, makes me to suspect that only a few buses with special groups of 
people were allowed to go through, not just ordinary buses for tourists.}. 

After the police inspection took place (highly unpleasant to say the least, with all of us lining up 
outside the bus) a local tourist guide came inside the bus and we entered the marvelous Unter den Linden 
(`Below the Linden Trees') Avenue, with four rows of beautiful lime trees. The high TV/radio tower
in Alexander Platz was also very impressive. After visiting various street monuments with political 
motives (where, again, we were alone) and the Pergamon museum, we came back at the 
Brandenburger Tor, where the police inspection was even more unpleasant than before, eagerly 
searching for people hidden in all imaginable and unimaginable places inside the bus. When we were 
lining up with our passports, the female officer in charge, who looked very proud of herself,
said to an Spanish student whose name was Fidel:  Fidel, ein wunderbare Name (a wonderful name) 

The day after, we had the afternoon free and I decided to come back to the East. I took the subway 
equipped with my passport, ready to cross the border at Friedrichstrasse.  At this station I encountered 
a large queue of people waiting underground: the access was in the interior of the station, not in the 
street. I had the impression that people there looked sad, sort of depressed, including the police officers. 
I think the grey and brownish colors at the place somehow contributed to the unpleasant feeling. In addition, 
everybody was forced to buy some money: 8 Marks minimum (I was lucky because two months afterwards this `fee' 
became 20 Marks). Apparently, 
the depressive mood of most people around was contagious since, once I finally reached the surface,
the adventure mood was completely gone. So I just walked a few hundreds meters until I found an 
appropriate caf\'e to stay for one hour or so. The caf\'e was quite spacious and well decorated. I ordered
Coca-Cola (yes, Coca-Cola in the communist world!). Back at the border the officers were asking
everybody if we had any money left ...... I secretely kept two coins.

\vskip .5in

\noindent
{\bf The Russians at MIT, 1988-1989}

\vskip .5in

After that interesting, although depressing, experience in Berlin I was even more curious than
before to come back to the communist world eleven years later. My last two years at MIT, in addition, were 
sort of illuminating in this respect as two Russian physicists came to stay a few months at the 
CTP in their way to establish themselves in the USA. 

The first Russian physicist to arrive at the CTP, around March 1988, was Mikhail 
Bershadsky (Michael, for everybody there).  The day he arrived I was giving a talk when 
he showed up in the seminar room looking disoriented and space-out. Single child from a Jewish family 
from Moscow, his parents had applied for fifteen years or so to be allowed to leave the Soviet Union. 
They finally got permission and planned to stay in USA for good. Michael was a great young guy
who had written a couple of successful papers in String Theory, although he still had to get his Ph. D. 
At the CTP he was very interactive, able to discuss about all kinds of matters (mainly String 
Theory and Mathematics). It was really convenient to have him around, as he was most helpful and motivating. 
He was also very outgoing and communicative with respect to social matters. As a matter of fact, I learned 
from him many stories about the daily life and difficulties in the Soviet Union, including the official (although 
hidden) discrimination against Jews. For example, he was told since he was a child  that he would never 
be admitted at the State University in Moscow, no matter how bright he would perform at school. On the 
other hand, Michael together with Alex, a Jewish student of Barton Zwiebach, also explained to me some 
of the eccentricities (in their own opinion) of the orthodox and/or ultra-orthodox Jews: to separate the intake 
of meat from milk at least three hours, the shaving of women's head just to place a wig on top, not to press 
the buttons of the elevators during the Sabbath\footnote{It happened that Michael and Alex were dying laughing 
outside my office so I asked them what was going on. They simply were having fun scoffing at some rules 
followed by conservative and hyperconservative Jews. Then they explained a bit the situation to me. I was 
most impressed with the women's wigs and also
with the `buttons of the elevators' affair during the Sabbath: `some elevators are programmed 
so that they stop at every floor and, in case this is technically impossible, someone may be hired to 
press the buttons'. This shocking information had such an impact on me that I even wondered, seriously, 
whether these rules could have been Einstein's inspiration for his famous statements regarding human 
intelligence. For example \cite{quo}: `The difference between stupidity and genius is that genius has its 
limits'. (Notice that Einstein had a very low opinion about religions in general, and the Jewish religion in
particular: `an incarnation of the most childish superstitions', as he explained in a famous letter 
written to the philosopher Eric Gutkind in January 1954.)  Anyway, it is also true that many Europeans 
arrive at similar conclusions regarding human lack of intelligence after spending some time living in the USA, 
without the need to invoke wigs or elevators. It seems that USA is the country of the extremes: for the good 
but also for the bad.},  etc. After the summer Michael Bershadsky moved to Princeton to work on his Ph.D. 
He did not have much luck in the following years, in the sense that he did not fulfill his very high 
expectations, and finally he left Physics heading to finances. 

The second Russian physicist arrived at the CTP around April 1989 to stay for a couple of months. He was 
Alexander Polyakov. Knowing he was coming, I was very moved, really enthusiastic. He was certainly one of my 
heroes, very high in the list\footnote{The fact that I had never see him before also contributed to his high 
position in my heroes list. This doesn't mean that he was ugly, but rather that heroes without an image tend to 
grow out of scale.}, 
and those days I was reading his recent book \cite{Poly}. In the CTP he was treated also 
like a hero. For example, Roman Jackiw offered him his spacious office. He started giving lectures from the 
very beginning, two or three per week. They were mainly about the stuff in his book
and from time to time he was making jokes of political connotations (about fences in the physical systems, etc.)
It was a pity that he never joined the group for lunch in the CTP common room, 
although we could see him often in the corridor in the way to his office, usually wearing very short pants
that looked almost like underwear. 
The first month or so he behaved quite correctly towards us, 
answering kindly all questions during his lectures, etc. Then he had the bizarre initiative of organizing 
one of his seminars at MIT on Sunday morning. I attended that seminar, not very enthusiastically, 
wondering why he was doing this to us. The mathematician Isadore Singer, who was very elegant 
wearing a suit (nobody else would do this at the CTP), introduced the talk smiling: 
`Well, it seems that Sasha likes to give talks on Sundays....'. A couple of months later, in June 1989, 
Sasha Polyakov left MIT and started his permanent position in Princeton University. I think at that time
he had moved several positions down in my heroes list.

One month later I met him again. It happened that I was invited to visit the IAS (Institute for Advanced 
Study) in Princeton for two weeks in July\footnote{I thank very much, again, Chiara Nappi and Edward Witten for 
that invitation. The two weeks at the IAS were really great and, in addition, I was allowed to keep the original 
key of the back door of Einstein's house, that I saved in its way to the garbage truck. Frank Wilczek, who was 
on vacations, was living in the upper floor of the house while the lower floor still had many 
items, like furniture, from Einstein. When I went there to have a look through 
the windows I noticed that the old door at the garden had been replaced but was still there lying 
on the wall with the key inside, so.....Back in Boston, Cumrun Vafa (my extra-official supervisor) was not 
amused when I showed him the Einstein's key:  `Look, what you have done is, precisely, what Einstein 
didn't want people to do!'}. Ed Witten had lunch every day with the IAS group, so there were plenty 
of opportunities to talk to him. In my first day there he sat in front of me at the table and asked me a couple 
of long questions about my recent work \cite{Bea} on non-abelian orbifolds and the Riemann-Hilbert 
problem\footnote{Cumrun Vafa proposed me to investigate the construction of vertex operators on non-abelian 
orbifolds using the Operator Formalism in 1988. 
Shortly afterwards Michael Bershadsky told me that this issue was probably related to the 
Riemann-Hilbert problem, and he was right.}. I guess this was his way to say hello to newcomers. 
The trouble was not that I didn't know the answers, but rather that I couldn't understand the questions. His
mathematical level was simply too high for me. So, I was relieved to know that in the
IAS no talks are scheduled in July. Otherwise, I was told, they would have invited me to present 
that paper. (As a matter of fact, I had given such a talk in June at Harvard and everything went fine, 
spite the fact that Paul Ginsparg was in the audience. Cumrun Vafa even said that it was a
very nice talk, but still I suspect it was better not to give it at the IAS.)

One Friday late in the afternoon, the Institute was very quiet. When I was about to leave I heard some 
people speaking just outside. When I arrived at the entrance door I wished I had a camera to capture 
the view forever. The two giants of Theoretical Physics from the two most powerful countries in the planet 
were just there talking to each other. They finished the conversation right at the moment 
I was getting out. Ed Witten went inside the Institute while Sasha Polyakov looked at me smiling:
 `What are you doing here? Are you coming to the Summer Workshop in Aspen?' No, I replied,
 `I am preparing my coming back to Europe in September. I will stay two years at CERN'. Then we were 
talking for a couple of minutes more and said goodbye...... 
apparently forever because I never talked to him again, although I saw him in some conferences
in Europe, but always at quite a distance\footnote{In any case, he is not easy to approach. For example,
in `Strings 1997' in Amsterdam he only attended his own session, and didn't join any social events. I also
heard that he didn't show up in a conference organized in his honor by his colleagues of Landau Institute.}.

\vskip .5in

\newpage

\noindent
{\bf Travelling to the USSR/CCCP, 1990}

\vskip .5in

After the Russian experience in MIT, not long ago, in October 1990 I was going to have the 
opportunity to encounter `the real thing', so I was very excited with the idea of visiting ITEP in 
Moscow. I received by telex the formal invitation, 
in order to get the visa, because the air mail with the Soviet Union worked very poorly (the e-mail 
not at all, but this situation improved soon afterwards). In no time, all the Russians 
around in the Theory Division  had heard about my trip to Moscow, so I received several useful 
advises for `survival purposes'. Since I arrived at CERN I had a few conversations with several Russian 
colleagues (this is why I got the invitation to visit ITEP in the first place). I especially remember talking 
to Ian Kogan, who was a very kind and warm person, really charming. He was at CERN for a few 
months and told me about his plans for the future. He wanted to find a permanent job, in Europe 
preferably, and never come back to Russia. He said that, in spite of the opening and reforms promoted 
by Mikhail Gorbachev (the Perestroika and Glasnost), the Soviet Union would still need 20 years more to 
become a place comparable to the West, with respect to living standards, and he did not want to wait 20 years 
to live a decent 
life\footnote{After working very hard for a few years, Ian Kogan got a permanent position in Oxford in 1994. 
Unfortunately, he continued working too hard afterwards and he died of a heart attack in Trieste, in 2003, 
at the age of 45.}.

One or two days before my departure to Moscow (October 7th) John Bell's funeral took place, the first 
funeral I attended where lunch was offered, and probably one of the rare occasions where John Ellis 
was wearing a suit. Mary was there talking to everybody, her calmness and fortitude were really impressive.

Alexander Turbiner drove me to the airport together with 6 Kg of packages for colleagues in Moscow.
Just before we left my apartment he said: `Now you are ready, so you must sit down again, close your 
eyes and think carefully to make sure you don't forget anything. This is a Russian habit'. So I did it.
Once in Moscow, after one or two hours delay, Andrei Losev was waiting for me at the airport. 
I had never met him before (I think he had never traveled abroad), and his English was very good.
The first thing he did was to give me a plane ticket and to report on the activities and instructions 
for the next few days. For a moment I felt I was more a spy than a scientist: 
`In two days you will fly to Odessa, at the Black See in Ukraine. I will accompany you'\footnote{Several 
years later I felt I was a spy again: I was spending a couple of months at CERN when an Israeli colleague
asked me if I could introduce him to a Lebanese colleague during a party of the Theory Division that was 
going to take place in two days. I was glad to do it. As a matter of fact, I liked very much these two people, 
but I never understood why they needed my help to talk to each other (as a good spy, I asked no questions).}.
Most physicists had already left Moscow by train because a workshop was going to take place in a vacation 
resort in Chernomorka, near Odessa\footnote{Chernomorka is the Soviet name for Lustdorf - one of the 
German colonies in  south Ukraine - established after the invitation of Tsarine Ekaterina the Great. I thank 
Vladimir Rubtsov, one of the organizers, for this information.}. They needed two days to arrive there. Until 
then I had to stay inside my apartment, I was told, although a student would come the day after to bring 
my passport back (they needed it for the formalities in order to fly to Ukraine) and to take me for a walk. 

Andrei was very talkative and at the 
time we reached the apartment (more than one and a half hours later) he had more than enough time to 
explain to me all about the present and recent past of the Soviet Union, with special emphasis on the 
communist propaganda pretending that the USSR was the best country in the world (that Andrei didn't believe). 
The ITEP apartment was really good and spacious. They had even provided some food in and out of the 
refrigerator (I brought some too in my suitcase, as advised). Andrei wrote for me the names and pronunciations 
of some groceries I might be interested to buy: milk, bread, yogurt,...  Surprisingly, in between 
the shopping list, he also started making comments and asking questions about my paper on non-abelian 
orbifolds \cite{Bea}. I wonder whether Andrei was chosen to accompany me because he was interested 
in some of my work or, rather, he read some of my work because he was chosen to accompany me.  Then we
went to a grocery store nearby and I bought some items. 

The day after, the student brought my passport back and took me for a walk, as promised. The `chosen 
one' for this important mission was Anton Gerasimov. He was very talkative too, like Andrei, but his
perspective on life and the USSR was quite different. He was very idealistic and probably believed in
communism as a great thing (although he didn't say that explicitly). For example, he argued that many 
people living in the USSR were far more free than many people living in the West, because freedom 
has to do mainly with the state of mind, etc.... Of course there was some truth in his reasoning. 
Everybody will agree that the ability to think freely of many people in this planet has shrunk to zero, 
essentially, due to ideological brainwash. 

The next day Andrei Losev picked me up and we went for a walk, then
we had lunch in a local restaurant - the meal consisted of two tasty dishes and dessert, like in Spain - 
and finally we went by taxi to the domestic airport. He was quite nervous because he was responsible for 
my safety. When we entered the plane we were separated from each other: the foreign
passengers were placed at the front whereas the Russians were placed behind, with a little curtain 
in between the two zones (like the separation between tourist class and business class in western planes). 
I went to the toilet and found three or four people sitting on the floor, with their baggage, just there.

Once we arrived at Odessa, around midnight,  Andrei realized nobody was waiting for us at the airport. 
Moreover, he didn't
know the address of the resort we were supposed to go, he only had some vague idea about its localization.
He called some people by phone, but he couldn't find much information. He was told that the trains were 
delayed for many hours, this was the reason why nobody showed up at the airport. Fortunately I had dollars,
so we took a taxi in order to approach Chernomorka. It was very late at night and a very polite, well dressed 
gentleman joined us. After more than one hour we arrived at his house and he invited us to spend the night
there since the taxi driver had no idea about the resort. It was around half past two, so we decided to
accept the invitation. He woke his mother up, who was extremely nice smiling at us all the time. It was a 
very big country house, and they put us in a very big living room with several bed-like objects. In the
morning the mother prepared breakfast for us. They didn't want any payment, so I slid carefully a 
20 dollar bill in one of her pockets. Andrei finally contacted someone who knew the exact address of the
resort, it was at a short distance from the place where we were. 

The resort was at the beach and the Black See was deep blue instead. It was a fantastic sunny day. 
Lots of physicists were already there, from everywhere in the USSR. There were lots of mathematicians 
too, they were having a parallel workshop as well. My host, Aliosha Morozov\footnote{Some nicknames
that Russian people use for the most common names are: Sasha for Alexander, Aliosha for Alexei, Dima for
Dimitri, Natasha for Natalia, Volodya for Vladimir, Boria for Boris and Irina for Ira. As a result, Belavin, 
Polyakov and Zamolodchikov amount to Sasha, Sasha and Sasha.}, was there with his wife 
Julia and baby Andrey. They looked a very happy family. Many other physicists had brought their families
too, there were many wives and some children who spent most time in the beach. About husbands, I
don't know if there were any at all because very few physicists were women, like it happens in north and 
central European countries and in the USA\footnote{In mediterranean European countries there were (and are) 
far more women in Physics and in scientific carriers than in the rest of Europe and USA. The reason could 
be that in these countries science is considered of interest for everybody, not especially for men, and 
in addition grandparents get very much involved in looking after the children.}.  

Since most physicists there had never traveled abroad I hardly knew a couple of them. Surprisingly, I felt very
much at ease among these people, as if they were  from Spain\footnote{There are many stories from Spanish
veterans who engaged to fight communism during the Second World War and ended up in Russia: they felt
very much at home among Russian people as if they were Spanish.}. So, I was socializing well and I didn't 
need any help, except during the talks: all of them in Russian. So, from time to time somebody there would 
explain me what was going on. Anyway, the sunny beach was also an option..... although I didn't have a 
swimsuit, of course!  The way they denoted
the meals was a bit confusing: everybody in the USSR had learned that the English word for lunch was 
`dinner', whereas dinner was `supper' (which is also correct, of course). After the dinners (i.e; the suppers)
a bunch of physicists from Moscow, mainly from ITEP, used to meet in one of the rooms to have fun: playing 
guitar, singing, drinking heavy stuff, eating snacks,.... I don't remember whether they invited me to participate 
or I went by my own, guided by the noise, but the fact is that I ended up in that room from the very first night. 
Like many Spanish compatriots I also played guitar..... 

I think it was my first day there, or the second one, when I met Vladimir 
Fock, grandson of the great physicist with the same name. He was doing his Ph.D. in ITEP under the 
supervision of Ian Kogan. We got along quite well from the very beginning and he offered to accompany me
to visit Leningrad (St. Petersburg nowadays) after the workshop. I accepted gladly, of course. 

In the morning of my second day in the resort I had a surprise. The gentleman who hosted Andrei and me,
two nights before, came to visit and brought me a book about Odessa. He was very grateful because 
I gave 20 dollars to his mother (at that moment a dollar was 6 rubles worth but in a few 
days it was going to be 20). 

One evening during dinner the light went off. This was not a major problem as everybody there was equipped
with candles, fully prepared for this eventuality.  I got a candle too (that I still keep as a souvenir).
Later at night we gathered in `the room' bringing our candles again, for in case. Some people 
there were a bit worried about me being scared. I assured them that I was not scared at all. I was aware
that `being from the West' was regarded as being more delicate and fragile and also more individualistic 
and demanding because we were used to lead a more comfortable life in the material sense (no shortage
of daily products, no queues for shopping, more cars, better salaries,...).

The last day of the workshop I was supposed to give a talk. Then someone told me: `we are sorry, you 
cannot give your talk because the mathematicians have finished their workshop ahead of schedule and
they have brought the blackboard along, we borrowed it from them'. 

Andrei Losev accompanied me again in my flight back to Moscow, whereas all other colleagues were 
traveling by train. In Moscow Andrei took me to a couple of walks, including the Red Square and 
Lenin's Mausoleum, where he continued his teaching of current and past political situation. By then
I had already understood that Russians (or perhaps Soviet people in general) didn't like Mikhail
Gorbachev at all, and they very much preferred Boris Yeltsin. Andrei also brought me to another 
local restaurant for lunch, and showed me a couple of shops. I bought lots of vinyl records, from 
both classic music and pop music (Beatles, Deep Purple,....), with the musicians names and 
everything else written in Cyrillic. His mother,
Natalia, invited me for lunch at home. She was a very beautiful, energetic and nice woman. I was really 
impressed by her. In addition I heard she was architect and had built some important buildings in Moscow.
Andrei's wife was also there although she didn't talk much. 

When Volodya Fock arrived from Chernomorka it was his turn to take care of me, so
he arranged things so that we could travel to Leningrad. We took a night train and we were hosted 
by a female friend of his mother: the renowned mathematician Olga Ladyzhenskaya who was the leader of the
`Leningrad School of Partial Differential Equations'. She worked in the Steklov Institute and was member of 
the Russian Academy of Sciences.  When we arrived to her flat Olga received us with very low voice 
saying: `Look, my niece is here in the living room. She 
works for the government and she is not allowed to have any contact with foreign people. So, please, rush
through the living room, enter the corridor, and take the two rooms at the end'. So, we followed the 
instructions and ran stealthily while the niece was looking through the window giving us the back. Then
Olga prepared some breakfast for us (this was the first time I had pasta for breakfast) and after that
we left the place and went for a walk. 

Leningrad was very beautiful, and the two or three days we were there we visited the 
Hermitage museum and the Steklov Institute, where some seminars were going on. One of 
those days we couldn't find any place to have some lunch (we tried in a couple of local restaurants 
but they were full). As a result we ended up in one of those restaurants where only dollars 
were accepted. At the entrance there were two tall security guards who looked mafia-related guys.
The interior was not much better: it took very long until they served us some simple dishes and the
soup cups were not filled to standard quantities, but to quite less. Volodya asked me: is it customary 
in the West to fill the soup cups so little? No, I replied. Anyway, these people did not look well at
all, so I decided not to complain and to eat our meal and leave the place as soon as possible.
We bought some red roses for Olga, Volodya told me the number should be odd because in
Russia an even number of flowers was believed to bring bad luck. Olga, in her turn, bought a
nice peanut cake for us (this was probably a luxury taking into account the economical situation).
She and Volodya were also very active in explaining the repression and excesses of the communism
to me, especially the horrors of the KGB. I remember Olga mentioning a well known scientist: `he 
simply disappeared, like many others, and nobody knew anything from him any more'.  

When we came back to Moscow Volodya accompanied me to several places of touristic interest: the
Kremlin\footnote{To visit the Kremlin there were different prices for Russians and for foreigners. 
So Volodya asked me to keep my mouth shut while he was buying tickets for two Russians..... },
the art market - where I bought many souvenirs - and the famous 16th-century Novodevichy 
Monastery with the equally famous Novodevichy Cemetery, where many important figures and 
celebrities are buried. There we visited the impressive grave of Lev Landau, among others. I guess
it was the religious atmosphere there what triggered a `profound' conversation. Volodya said
that he had `envy' of people who believed in God. My personal experience was exactly the
opposite, I replied, because when I was a young adolescent I was very unhappy and angry thinking 
why God allowed all the suffering and misery in this world for humans and animals\footnote{For example,
why God didn't create all animals and humans as vegetarians? This way no animal or human had to 
kill and hurt any other human and animal and most cruelty would be eradicated. About reproduction 
rates, the Almighty could slow down the reproduction rate of all species, so that they could 
live in harmony with each other. And how about horrible and very painful diseases (affecting often 
children)? Could anybody believe that these diseases had been designed by an intelligent Creator?}. Then, 
when I was about 14 years old I had the brilliant  idea that perhaps there was no Creator at all, therefore 
nobody to be blamed for all the `design failures' in this world and nobody to feel angry about. 
This idea brought me very much relief and peace of mind since.  

During one of our walks we passed near ITEP, although we didn't come inside. At the wall there was a
portrait of Ian Kogan. Volodya explained that he had been chosen `the worker of the month' or 
something similar, but Ian was surely not happy about it. This procedure I had seen before only in some 
supermarkets in the USA. 

Volodya's mother also invited me for lunch at home and her name was Natalia too! She was the 
daughter of Vladimir Fock and she was chemist. Volodya was her only kid and was living with her.    
When we were on our way to the flat, he explained that it was located
in a building of the so-called `vampire style' because its construction was ordered directly by Stalin.
The building was very high and they lived around floor 13, so they enjoyed magnificent views of Moscow.
Volodya said that Stalin respected Piotr Kapista - the most prominent Russian experimentalist,  
who discovered superfluidity - very much and, consequently, he could do things that nobody else could.
Fortunately, Kapista was a very good friend of Volodya's grandfather and used his influence to provide 
a good flat for Fock's family in Moscow and another one in Leningrad, although the family didn't keep that 
one anymore. This family was treated very well during all the hard years of communism and only now, in the
very last years, they were getting into financial trouble. 

Once in the flat with Natalia it became 
quite clear that the success and happiness of Volodya was the most important issue for her. She was 
a lovely person, very kind and affectionate. She prepared the typical soup `borsh', this was my first time 
having it, I think. During lunch we had an interesting conversation, as usual with Russian people. For 
example Volodya, who had never traveled abroad, explained: `Until now to travel to the West seemed to 
us more unlikely than traveling to the Moon'. The anecdote of the chocolate cake was also remarkable. Natalia 
started explaining that many years ago there was an abundance of food and most current items. Then, strangely,
some items started disappearing without notice. One day was the wine from Georgia, another day was this,
then was that, etc. Then Volodya continued: `There were two official cakes in the Soviet Union: the 
chocolate cake and another one. Then one day they also disappeared without any explanation. A friend
of mine saw the chocolate cake in Leningrad two years ago. It was sold at the entrance of a metro station.'

At a given moment Natalia looked at me and said: `You know, you are the second foreign person to enter 
in this flat. The first was Niels Bohr'. The surprise was enormous for me: I was the second after Niels Bohr  
in something! This was really impressive. Then Natalia explained that Niels Bohr invited her father to 
visit Denmark and she accompanied him. It was a wonderful trip. Later Niels Bohr paid
a visit to her father in Moscow, in 1961. Bohr didn't live much longer as he died in 1962. 

The day after this lunch I came back to CERN. Two months later, in December, I took part in the Christmas
theater play of the Theory Division at CERN. I played two roles. First I was a piece of Gruy\`ere cheese
that had to melt with some Emmental cheese in order to make a Swiss fondue, and next I was a 
neurotic scientist fighting with another even more neurotic scientist (the reader may guess who was chosen, 
by John Ellis, to play the roles of the Emmental cheese and the scientist fighting with me). This was 
my last unforgettable experience in the annus mirabilis 1990: I had never before, and I have never again,
taken part in a theater play.

The USSR/CCCP started collapsing in March 1991, it continued doing so during the summer, with an abortive 
attempt of coup d'\`etat  in August, and finally it was officially dissolved in December 1991. 
The Commonwealth of Independent States (CIS) was established in its place.

In April 1992 I came back to Moscow, now part of the Russian Federation. This time I was invited by 
the Lebedev Institute. Natalia Fock hosted me at home and Volodya picked me up at the airport. 
I was installed in a big room with a blackboard and an equally big sofa. Natalia was living 
alone, with Cotangent the cat, since her son was living somewhere else with his girlfriend. 
Unlike my previous experience in the USSR, this time I was leading
a normal daily life in Moscow, like the Russians themselves. This reminded me very much of the stories 
told by my parents about the post-war time in Madrid after the Spanish Civil War (1936-1939): 
hardship, hardship and hardship, in every sense. In addition,
it was not easy at all to give money to Natalia. She didn't want to charge me anything for my stay there,
she wanted me to feel invited in her place. So, the only way I figured out was handing over some money to
her so that she could buy some presents I wanted to give them (fortunately presents were allowed). I also
asked her to let me know in case they needed financial support, so that I could send them some money. I 
never heard from her about this, perhaps because Volodya started travelling abroad very soon (Uppsala 
University in Sweden was his first destiny). When I returned to CERN, only two and a half weeks later, I said to 
my husband: the trip took 5 hours and 50 years: I really felt I was coming from the past, like from a time-machine.

\vskip 1in

\vskip .2in
\noindent
{\bf Acknowledgements:}
\vskip .2in
\noindent
I am very grateful to Alexei Morozov for the invitation to visit ITEP in 1990 - although I never visited ITEP - and 
for all the nice gestures towards me. He provided a great  ITEP apartment in Moscow, a plane ticket to
Odessa, a sunny week at the beach - although I didn't have a swimsuit - and two fantastic bodyguards. 
I thank very much the latter - Andrei Losev and Vladimir Fock - for taking care so well of me, and I also thank 
their mothers - the two Natalias - and Olga Ladyzhenskaya for their hospitality and nice meals.
This work has been partially supported by funding of the Spanish Ministerio de Ciencia e Innovaci\'on, 
Research Project FPA2008-02968, and by the Project CONSOLIDER-INGENIO 2010, Program CPAN (CSD2007-00042). 
%\newpage

\bibliography{REFS}
\bibliographystyle{lennaert}

\end{document}